\documentclass[preprint,aps,12pt]{revtex4}

\usepackage{graphicx}% Include figure files
\usepackage{dcolumn}% Align table columns on decimal point
\usepackage{bm}% bold math
\usepackage{amsfonts,amsmath,amsthm,graphicx}

\begin{document}

%\preprint{APS/123-QED}

\title{Operating an atom interferometer beyond its linear range
%Sensitivity of an atomic gravimeter without vibration isolation
%Operating an atom interferometer beyond its linear range
%the limit imposed by high vibration noise
}% Force line breaks with \\

\author{S. Merlet}
\author{J. Le Gou\"{e}t}
\author{Q. Bodart}
% \altaffiliation[Also at ]{Physics Department, XYZ University.}%Lines break automatically or can be forced with \\

\author{A. Clairon}%
\author{A. Landragin}%
\author{F. Pereira Dos Santos}%
 \email{franck.pereira@obspm.fr}

\affiliation{
LNE-SYRTE, CNRS UMR 8630, UPMC, Observatoire de Paris\\
61 avenue de l'Observatoire, 75014 Paris, France}

\author{P. Rouchon}%
\affiliation{%
Mines ParisTech, Centre Automatique et Syst\`emes\\
60, bd. Saint-Michel, 75272 Paris Cedex 06, France
}%

\date{\today}% It is always \today, today,
             %  but any date may be explicitly specified

\begin{abstract}
In this paper, we show that an atom interferometer inertial sensor, when associated to the auxiliary measurement of external vibrations, can be operated beyond its linear range and still keep a high acceleration sensitivity. We propose and compare two measurement procedures (fringe fitting and nonlinear lock) that can be used to extract the mean phase of the interferometer when the interferometer phase fluctuations exceed $2\pi$. Despite operating in the urban environment of inner Paris without any vibration isolation, the use of a low noise seismometer for the measurement of ground vibrations allows our atom gravimeter to reach at night a sensitivity as good as $5.5\times10^{-8}$g at 1 s. Robustness of the measurement to large vibration noise is also demonstrated by the ability of our gravimeter to operate during an earthquake with excellent sensitivity. Our high repetition rate allows for recovering the true low frequency seismic vibrations, ensuring proper averaging. Such techniques open new perspectives for applications in other fields, such as navigation and geophysics.

\end{abstract}

\pacs{Valid PACS appear here}% PACS, the Physics and Astronomy
                             % Classification Scheme.
%\keywords{Suggested keywords}%Use showkeys class option if keyword
                              %display desired

\maketitle

\section{Introduction}

Atom interferometers \cite{Borde89} are used to develop highly sensitive inertial sensors,
which compete with state of the art ``classical'' instruments~\cite{Niebauer95}. Applications of such interferometers cover numerous fields, from fundamental physics \cite{Fixler07,Lamporesi08,hsurm1,hsurm2,vigue} to navigation and geophysics. For instance, transportable devices are being developed with foreseen applications in the fields of navigation, gravity field mapping, detection of underground structures ...

In most of these experiments, atomic waves are separated and recombined using two-photon transitions, induced by a pair of counterpropagating lasers. The inertial force is then derived from the measurement of the relative displacement of free-falling atoms with respect to the lasers equiphase, which provide a precise ruler. As the inertial phase
shift scales quadratically with the interrogation time, very high sensitivies can be reached using cold atoms along parabolic
trajectories \cite{Kasevich91,Canuel06}, provided that the experiments are carefully shielded from ground vibrations. In the usual geometry where the laser beams are retroreflected on a mirror, the position of this mirror sets the position of the lasers equiphase, so that only this "reference" optical element is to be shielded from ground vibrations. Such an isolation can be realized either with an active stabilization scheme, using a long period superspring \cite{Niebauer95,Peters01,Hensley99}, or by using a passive isolation platform \cite{LeGouet08}. For instance, the use of a superspring allowed increasing the interaction time up to 800~ms and reaching a best short term sensitivity to acceleration of $8\times 10^{-8}\textrm{m.s}^{-2}$ at 1~s \cite{Mueller08}. An alternative technique, which we study in this article, doesn't require any vibration isolation, but exploits an independent measurement of ground vibrations, realized by a low noise accelerometer. A technique based on the same principle has already been used with a "classical" corner cube gravimeter \cite{Canuteson97,Brown01}, and allowed improving its sensitivity by a factor 7 \cite{Brown01}.

In this article, we investigate the limits to the sensitivity of an atomic gravimeter when operating without vibration isolation. This transportable gravimeter is developed within the frame of the watt balance project led by the Laboratoire National de M\'{e}trologie et d'Essais (LNE)~\cite{Geneves05,Merlet08}. We first briefly describe our experimental setup, and recall the usual procedures for measuring the mean phase of the interferometer. We then introduce and compare two measurement schemes (finge fitting and nonlinear lock) that allow operating the sensor in the presence of large vibration noise, and show how phase measurements can be performed even though the interferometer phase noise amplitude exceeds $2\pi$. These schemes, which use an independent measurement of vibration noise with a low noise seismometer, allow to reach good sensitivities without vibration isolation. In particular, we reach a sensitivity as good as $5.5\times10^{-8}$g at 1 s during night measurements, in the urban environment of inner Paris. Finally, the robustness of these measurement schemes versus changes in the vibration noise is illustrated by the capability of our instrument to operate and measure large ground accelerations induced by an earthquake.

\section{Limits due to vibration noise in a conventional setup}

\subsection{Experimental setup}
The experimental setup, which we briefly recall here, has been described in detail in \cite{LeGouet08,Cheinet06}. About $10^7$
$^{87}$Rb atoms are first loaded in a 3D-MOT within 50 ms, and further cooled down to $2.5~\mu\textrm{K}$ before being dropped in free fall. Before creating the interferometer, a narrow vertical velocity
distribution of width about 1 cm/s is selected in the $\left|F=1,
m_F=0\right\rangle$ state, using several microwave and
optical Raman pulses.

The interferometer is then created using Raman transitions \cite{Kasevich91} between the two hyperfine levels $F=1$ and $F=2$ of the $^5S_{1/2}$ ground state, which are induced by two vertical and counterpropagating laser beams of frequencies $\omega_1,\omega_2$ and wavevectors $\vec{k}_{1},\vec{k}_{2}$. A sequence of three Raman pulses
($\pi/2-\pi-\pi/2$) allows to split, redirect and recombine the
atomic wave packets. The relationship between external
and internal state \cite{Borde89} allows to measure the interferometer phase shift
out of a fluorescence measurement of the
populations of each of the two states. At the output of
the interferometer, the transition probability $P$ from one hyperfine
state to the other is given by $P=a +
b\cos\Delta\Phi$, where $2b$ is the interferometer contrast,
and $\Delta\Phi$, the difference of the atomic phases accumulated
along the two paths, is given by $\Delta\Phi=-\vec{k}_\text{eff} \cdot
\vec{g}T^{2}$ \cite{Borde01}. Here
$\vec{k}_\text{eff}=\vec{k}_{1}-\vec{k}_{2}$ is the effective wave
vector (with $|\vec{k}_\text{eff}|=k_1+k_2$ for counter-propagating
beams), $T$ is the time interval between two consecutive
pulses and $g$ is the gravity acceleration.

The Raman light sources are two extended cavity diode lasers based
on the design of \cite{Baillard06}, which are amplified by two
independent tapered amplifiers. Their frequency difference, which is
phase locked onto a low phase noise microwave reference source, is swept according to $(\omega_2-\omega_1)(t)=(\omega_2-\omega_1)(0)+\alpha t$ in order to compensate for the gravity induced Doppler shift. This adds $\alpha T^2$ to the interferometer phase shift, which eventually cancels it for a perfect Doppler compensation, for which $\alpha_0=\vec{k}_\text{eff} \cdot
\vec{g}$.

\subsection{Conventional measurement procedures}
\label{conv}
Maximal sensitivity to phase fluctuations is achieved when operating the interferometer at mid fringe, which corresponds to $\Delta\Phi=\pm\pi/2$ . In this case though, variations in the offset $a$ can be interpreted as fluctuations of the interferometer phase. A standard technique \cite{Peters01} consists then in recording a full fringe, by measuring the transition probability as a function of a controlled phase shift induced on the interferometer. Fitting this fringe then allows measuring $g$. This technique degrades the short term sensitivity as measurements performed at the top or bottom of the fringes are not sensitive to phase fluctuations. An alternative way consists in using a method inspired by microwave atomic clocks. The phase is modulated by $\pm \pi/2$ so that the measurement is always performed at mid fringe, alternatively to the right and to the left side of the central fringe. From two consecutive measurements $P_i$ and $P_{i+1}$, the phase error can be estimated. In practice, a correction $G\times(P_i-P_{i+1})$ is added at each cycle to $\alpha$, in order to stir the chirp rate onto the central fringe. This realizes an integrator, whose time constant can be set to a few cycles by adjusting the gain $G$. This locking technique has the advantage of rejecting offset and contrast fluctuations, while preserving maximal sensitivity to phase fluctuations.

\subsection{Influence of vibration noise}
\label{vibnoise}
In the case where the duration of the Raman pulses can be neglected, the phase shift $\Delta\Phi$ induced by vibrations is given by
\begin{equation}
\label{deltaphi}
\Delta\Phi=k_\text{eff}(z_g(-T)-2z_g(0)+z_g(T))=k_\text{eff}\int^{-T}_{T}g_s(t)v_g(t)dt
\end{equation}
where $z_g$ and $v_g$ are the position and velocity of the experimental setup, and $g_s$ is the sensitivity function \cite{Cheinet08}, given by
\begin{equation}
\label{biggeq}
g_s(t)=\left\{
\begin{array}{lll}
 -1 &   &-T<t<0 \\
 1 &   &0<t<T\\
\end{array}
\right.
\end{equation}
The expected sensitivity of the interferometer to g fluctuations of the interferometer is then given by a weighted sum of the vibration noise at the harmonics of the cycling rate $f_{\rm{c}}$ \cite{Cheinet08}:
\begin{equation}
\label{sigmavib}
\sigma^{2}_{g}(\tau)={1\over \tau}\sum_{k=1}^{\infty}\left(\frac{\textrm{sin}(\pi k f_{\rm{c}} T)} {\pi k f_{\rm{c}} T}\right)^4
        S_{a}({2\pi k f_{\rm{c}}})
\end{equation}
where $\sigma_{g}(\tau)$ is the Allan standard deviation of acceleration fluctuations for an averaging time $\tau$, $S_{a}$ is the power spectral density of acceleration fluctuations.

\begin{figure}[h]
       \includegraphics[width=10 cm]{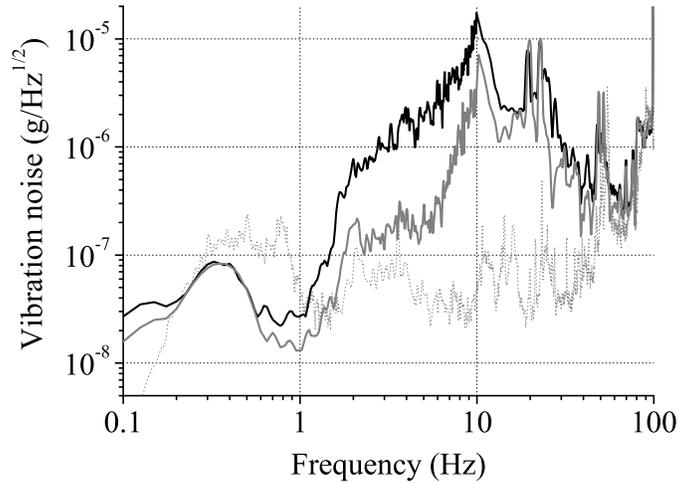}
    \caption{Amplitude spectral densities of vibration noise.
    The black (resp. grey) thick curve displays the vibration noise with the isolation platform down (OFF) at day time (resp. night time), while the dotted curve displays the vibration noise with the floating platform (ON) at day time.}
    \label{spectrevib}
\end{figure}

Figure \ref{spectrevib} displays the power spectral densities of vibrations, measured with a low noise seismometer (Guralp CMG-40T, response option 30s) on the platform which is either floating (ON) (day time), or put down (OFF) (day time and night time). In the case where the platform is OFF, the spectrum is similar to the spectrum measured directly on the ground. For our typical parameters, $2T=100$~ms and $f_c=3.8$~Hz, we calculate using eq.\ref{sigmavib} sensitivities at $\tau=1$~s of $2.9\times10^{-6}g$ during the day and $1.4\times10^{-6}g$ during the night with the platform OFF. With the platform ON, the sensitivity is expected to be $7.6\times10^{-8}g$.

\section{Vibration noise correction}

\subsection{Correlation between atomic and seismometer signals}

The signal of the seismometer can be used to determine the phase shift of the interferometer due to residual vibrations, as measured by the seismometer, $\phi_{vib}^S$, which is given by:
\begin{equation}
\label{deltaphi}
\phi_{vib}^S=k_\text{eff}\int^{-T}_{T}g_s(t)v_s(t)dt=k_{eff}K_s\int^{-T}_{T}g_s(t)U_s(t)dt
\end{equation}
where $U_s$ is the seismometer voltage (velocity) output and $K_s=400.2 V/(m.s^{-1})$ is the velocity output sensitivity of the seismometer.

\begin{figure}[h]
       \includegraphics[width=14 cm]{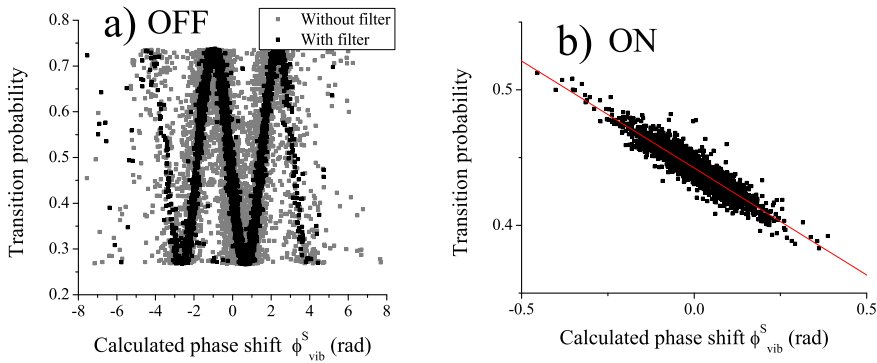}
    \caption{Correlation between the transition probability of the interferometer and the phase shift calculated from the seismometer data, for $2T=100$~ms a) The isolation platform is OFF. Grey points : without digital filter, black points : with digital filter. b) The isolation platform is ON. Black points : with digital filter. Line : fit to the data, with correlation factor of 0.94.}
    \label{correl}
\end{figure}

Figure \ref{correl} displays the measured transition probability as a function of $\phi_{vib}^S$, in the two cases of platform ON and OFF, for an interferometer time $2T=100$~ms. The noise is low enough in the ON case (fig \ref{correl}b)) for the interferometer to operate close to mid fringe, while in the OFF case (fig \ref{correl}a)), interferometer phase noise is larger than $2\pi$, and the interferometer signal jumps from one fringe to another. Figure \ref{correl} shows the good correlation between measured and calculated phase shifts. In the ON case, we find a correlation factor as high as 0.94.

\begin{figure}[h]
       \includegraphics[width=10 cm]{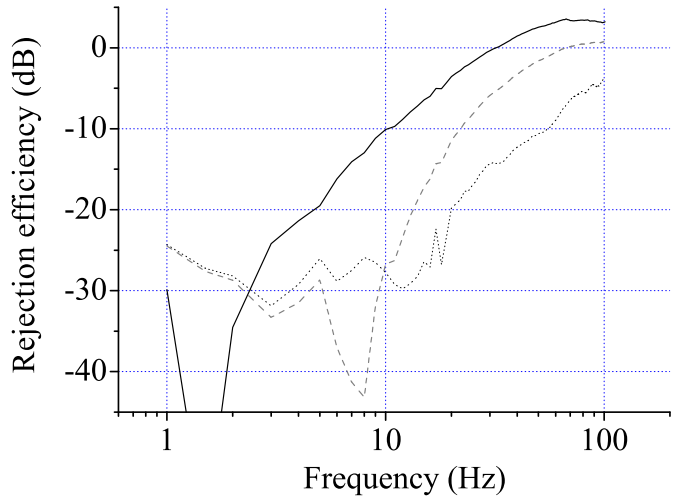}
    \caption{Efficiency of the vibration rejection as a function of frequency without any processing (black straight line), with digital filter (dashed line), with a compensation of a delay of 4.6 ms (dotted line). }
    \label{rejections}
\end{figure}

The calculated $\phi_{vib}^S$ can thus be used to improve significantly the sensitivity of the measurement, by applying a post-correction on the transition probability measured at mid fringe. This correlation is not perfect though due to the response function of the seismometer, which is not flat, and behaves as a low pass filter with a cut-off frequency of 50 Hz. This response function thus limits the efficiency of the vibration rejection. Figure \ref{rejections} diplays as a continuous black line the rejection efficiency as a function of frequency, which is calculated out of the seismometer transfer function.

\subsection{Digital filtering}

We implemented a numerical filtering of the seismometer signal to compensate for the phase lag of the seismometer at intermediate frequencies. The design of the filter is described in detail in \cite{LeGouet08}. It consists in the product of a recursive IIR (Infinite Impulse Response) filter, with corner frequencies $f_0$ and $f_1$, and a non-causal low-pass filter. The IIR filter compensates the phase shift of the seismometer signal and the non causal filter prevents the IIR filter from amplifying the intrinsic noise of the seismometer at high frequencies, without affecting the phase advance needed to improve the rejection.
The total transfer function of the filter is given by
\begin{equation}
F(f)=\frac{1+jf/f_0}{1+jf/f_1}\frac{1}{1+(f/f_c)^2}
\end{equation}
where $f_0, f_1$ and $f_c$  are then optimized in order to reach the best sensitivity.
This digital filtering improves significantly the rejection efficiency, as can be seen in figure \ref{rejections}, where it is displayed as a dashed line, for the frequencies $f_0=30$ Hz, $f_1=180$ Hz and $f_c=29$ Hz. Despite this increase in the rejection efficiency, the gain on the sensitivity, when implementing this filter in the ON mode, was limited to 25\% only \cite{LeGouet08}, which we attributed to excess noise of the seismometer arising from coupling between horizontal and vertical axes.

\subsection{Cross couplings}
\begin{figure}[h]
       \includegraphics[width=10 cm]{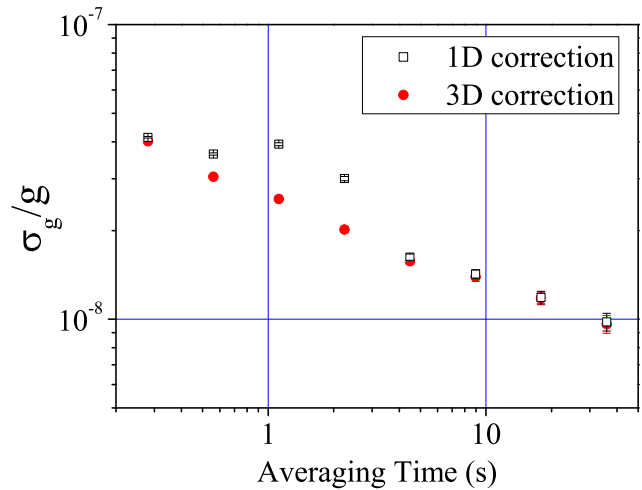}
    \caption{Sensitivity to $g$ with 1D and 3D corrections. The measurement was realized during the day, with a floating platform.}
    \label{corrhor}
\end{figure}

In order to detect these couplings, we recorded simultaneously the seismometer outputs along the three directions, calculated three corrections, one along each axis (only the vertical correction was numerically filtered though) and fitted the transition probability measured at mid fringe with a linear combination of the three corrections. The result of this fit showed couplings of 4\% and 5\% with the horizontal axes. We finally determined the influence of these couplings onto the sensitivity of the measurement, by comparing the Allan standard deviation of the phase fluctuations in the case where the correction is performed only with the vertical correction (1D), or with the optimal combination of the three (3D). The results are shown on figure \ref{corrhor}, where the sensitivity is expressed relatively to $g$.
Using the three corrections allows to remove a bump that appears when using the 1D correction. This indicates that horizontal vibration noise, as it also appears in the vertical seismometer signal, adds noise when performing a 1D correction.

\subsection{Efficiency of the filter without vibration isolation}

The digital filter is much more efficient in the OFF mode, as one can see on figure \ref{correl} a) where the noise on the interferometer fringes is significantly reduced when seismometer data are processed with the digital filter. In that case, the dominant contribution of the vibration noise onto the degradation of the sensitivity corresponds to frequencies around 10 Hz, for which the effect of the filter improves the rejection efficiency from 10 dB to about 30 dB \cite{LeGouet08}.

\subsection{Case of a pure delay}

We later noticed that the phase lag of the seismometer signal varies almost linearly with respect to frequency in the 1-100 Hz band, with a slope corresponding to a delay of about 5 ms. The phase shift of the seismometer can thus be compensated for, by simply shifting the acquisition of the seismometer data by this delay. We measured the correlation factor as a function of the delay, with the platform OFF, and found an optimal delay of 4.6 ms. The rejection efficiency for this optimal delay is displayed as a dotted line on figure \ref{rejections}. Surprisingly, we find a correlation similar to the optimal digital filter, despite a significantly different behavior of the rejection efficiency versus frequency.

\section{Measurements protocols}

\subsection{Standard procedures}

The standard measurement protocols described above need that phase fluctuations remain significantly smaller than $2\pi$. This requires to reduce the interferometer duration in the OFF mode to $2T \leq 20 $~ms. For $2T=20$~ms, the integrator scheme described above allows reaching sensitivities of $1\times10^{-5}g$ at 1~s when applying no correction to the measured transition probability, of $5\times10^{-6}g$ when correcting without filtering, and of $1.5\times10^{-6}g$ when correcting with digital filtering. The simple post-correction (without filter) thus improves the sensitivity by a factor 2, and the digital filter improves it further by a factor 3.5. Better performances are expected with large interrogation time for which the transfer function of the interferometer filters more efficiently high frequency vibration noise. In order to operate the interferometer with large interrogation times despite excess noise, we propose two alternative measurement procedures described in the following subsections. Both are based on the combination of measurements of the transition probability and of $\phi_{vib}^S$ by the seismometer. Though developed for the case of large vibration noise, these techniques can be extended to low vibration noise by adding a well controlled phase modulation.

\subsection{Fringe fitting}

The first technique simply consists in fitting fringes, as in \cite{Peters01}, except that here the phase of the interferometer is now scanned randomly by vibration noise. The signal displayed in figure \ref{correl} and obtained when plotting the transition probability versus $\phi_{vib}^S$, calculated with the digital filter, can be fitted by the function $P=a+b\cos(\eta \phi_{vib}^S+\delta\phi)$, where $a, b, \eta$ and $\delta\phi$ are free parameters. Due to the influence of the seismometer transfer function, $\eta$ will in general differ from 1. In practice, we operate the interferometer close to the central fringe, which corresponds to a small phase error $\delta\phi$. Every 20 points, we perform a fit of the signal and extract a value for the phase error $\delta\phi_m$. We then calculate the Allan standard deviation of the $\delta\phi_m$ in order to determine the sensitivity of the measurement. Note that this fitting procedure is not very efficient if the noise amplitude is significantly less than $2\pi$, because the interferometer signal remains close to the bottom of the central fringe. An additional and perfectly controlled phase modulation of $\pm \pi/2$ is thus applied in order to optimize the sensitivity of the interferometer to phase fluctuations. Moreover, the sensitivity improves by about 50\% when taking cross couplings of the seismometer into account, which can be realized by adjusting the data with a linear combination of the corrections along three directions $\Sigma \eta_j \phi_{vib,j}^S$, where $j={x,y,z}$ and $\phi_{vib,j}^S$ is the phase shift calculated out of the filtered seismometer data along axis $j$.

\subsection{Nonlinear lock}

The lock procedure described in \ref{conv} can be adapted in the case where the phase noise exceeds $2\pi$. Let's consider the measurement at cycle $i$ of the transition probability $P_i$
\begin{equation}
    P_i= a - b \cos((k_\text{eff} g - \alpha)T^2 + S_{i}) \\
    = a - b
   ( \cos e \cos S_{i}
  -  \sin e \sin S_{i})
\end{equation}
where $e=(k_\text{eff} g - \alpha)T^2$ is the phase error and $S_{i}$ is the phase shift induced by residual vibrations, estimated from the seismometer signal. We assume here that the phase error $e$ varies slowly, so that we can consider it as constant between three consecutive measurements.
Eliminating $a$ and $\cos e$  from the following three equations
\begin{align*}
    P_{i-1} &= a - b ( \cos S_{i-1} \cos e  -  \sin S_{i-1}  \sin e)
    \\
    P_i &= a - b ( \cos S_{i} \cos e  -  \sin S_{i}  \sin e)
    \\
    P_{i+1} &= a - b ( \cos S_{i+1} \cos e  -  \sin S_{i+1}  \sin e)
\end{align*}
gives
$$
b B_i \sin e = A_i
$$
with
\begin{align*}
    A_i &= (\cos S_{i+1}- \cos S_{i})(P_{i-1}-P_{i})
        - (\cos S_{i-1}- \cos S_{i})(P_{i+1}-P_{i})
    \\
    B_i &=
    (\cos S_{i+1}- \cos S_{i})(\sin S_{i-1} - \sin S_{i})
       - (\cos S_{i-1}- \cos S_{i})(\sin S_{i+1} - \sin S_{i})
\end{align*}

In order to stir the chirp rate onto the doppler shift rate, an iterative correction is applied on $\alpha$ according to
\begin{equation}\label{ordre1:eq}
    \alpha_{i+2} = \alpha_{i+1}+ K \frac{2 B_{i}}{1+B_{i}^2} A_{i}
\end{equation}
where $K$ is a positive gain. Here $\frac{2B_{i}}{1+B_{i}^2}$ is used as a pseudo inverse of $b B_{i}$ with $b\approx 1/2$, in order to prevent the correction from diverging when $B_{i}$ is close to zero. Choosing $K<1/T^2$ guarantees the stability of the servo loop.

\subsection{Adaptation  of the nonlinear lock}
\label{extension}
When phase fluctuations are significantly less than 1 radian, $Bi$ becomes much smaller than 1 (note that $Bi$ is null in the absence of vibration noise, which implies that the lock scheme doesn't work, as it is not able to stir the chirp rate), so that $\frac{B_{i}}{1+B_{i}^2}$ is not a good pseudo-inverse of $B_i$. This decreases the effective gain of the loop, which can be compensated for either by increasing $K$, or by replacing $\frac{B_{i}}{1+B_{i}^2}$ with $\frac{B_{i}}{\sigma_B^2+B_{i}^2}$, where $\sigma_B$ is the standard deviation of the $B_i$'s.

The scheme is then modified by adding extra phase shifts in order to increase the sensitivity to phase fluctuations. A simple phase modulation of $\pm \pi/2$, which implies that the interferometer operates alternatively at the right and left sides of the central fringe, is not sufficient, as in that case, $B_i$ is still null for null vibration noise. With a 3-phases modulation ($-\pi/2, 0, \pi/2$), $B_i=1$ for null vibration noise, and replacing $\frac{B_{i}}{1+B_{i}^2}$ with $\frac{B_{i}}{\sigma_B^2+B^2+B_{i}^2}$, with $B$ the mean of $B_i$'s, guarantees the full efficiency of the lock, whatever the amplitude of vibration noise.

The lock technique can be further modified to first determine and servo the vibration phase coefficients $\eta_j$. The phase of the interferometer is $e+S_{i}+\delta\phi_i$, where $\delta\phi_i$ is a controlled additional phase shift (alternatively $-\pi/2, 0, \pi/2$), and the vibration phase $S_{i}$ is (best approximated by) $\Sigma \eta_j \phi_{vib,j}^S$, where $j={x,y,z}$ and $\phi_{vib,j}^S$ is the phase shift calculated out of the seismometer data along axis $j$. At the i-th measurement, $S_{i}$ is calculated by $\sum_{j=1}^{3} \eta_{j,i} \phi_{vib,j,i}^S$, where $\eta_{j,i}=\eta_{j}-\delta\eta_{j,i}$. $P_i$ is thus given by
\begin{align*}
 P_i= a - b \cos(\delta\phi_i + \sum_{j=1}^{3} \eta_{j,i} \phi_{vib,j,i}^S +e+\sum_{j=1}^{3} \delta\eta_{j,i} \phi_{vib,j,i}^S) \\
 P_i=  a - b(\cos S_{i}
  -  (e+\sum_{j=1}^{3} \delta\eta_{j,i} \phi_{vib,j,i}^S)\sin S_{i})\\
\end{align*}
where $S_{i}=\delta\phi_i + \sum_{j=1}^{3} \eta_{j,i} \phi_{vib,j,i}^S$.

Generalizing the algebra above, one gets
\begin{equation}
b (B_i e+\sum_{j=1}^{3} C_{j,i}\delta\eta_{j,i}) = A_i
\end{equation}
where
\begin{align*}
   C_{j,i} =
    (\cos S_{i+1}- \cos S_{i})(\phi_{vib,j,i-1}^S\sin S_{i-1} - \phi_{vib,j,i}^S\sin S_{i})\\
       - (\cos S_{i-1}- \cos S_{i})(\phi_{vib,j,i+1}^S\sin S_{i+1} - \phi_{vib,j,i}^S\sin S_{i})
\end{align*}
Chirp rates and vibration phase coefficients are then corrected according to
\begin{align*}
    \alpha_{i+2} = \alpha_{i+1}+ K \frac{B_{i}}{\sigma_B^2+B^2+B_{i}^2} A_{i}\\
    \eta_{j,i+2} = \eta_{j,i+1}+L_j \frac{C_{j,i}}{\sigma C_j^2+C_{j,i}^2} A_{i}
\end{align*}
where $L_j$ is the gain for direction $j$.
Such nonlinear feedback and estimation algorithms are inspired from Lyapounov stability theory, for the main loop given by eq. \ref{ordre1:eq}, and  adaptive techniques, for the estimation of parameters $\eta_j$ (see~\cite{slotine-li-book} for a tutorial presentation of such techniques and~\cite{khalil-book} for a more advanced one).

\begin{figure}[h]
       \includegraphics[width=15 cm]{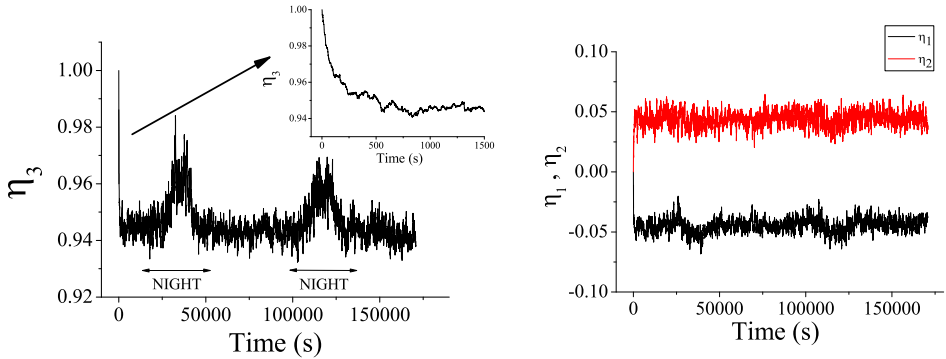}
    \caption{Evolution of the vibration phase coefficients, during a measurement realized using the nonlinear lock scheme, with initial settings $\eta_{j,0} = (0,0,1)$. The graph on the left (resp. right) displays the vertical (resp. horizontal) phase coefficient(s).}
    \label{coeff}
\end{figure}

Figure \ref{coeff} displays the evolution of the vibration phase coefficients during a two day measurement. The time constant of the lock is about 200 s (see inset). Note that the vertical phase coefficient $\eta_3$ differs significantly from 1, and is different at day and night times, which can be attributed to a change in the vibration noise PSD. Moreover, the lock converges towards horizontal phase coefficients of about 5\%, in agreement with the values previously determined with the fit.

\subsection{Comparison of the two techniques}

\begin{figure}[h]
       \includegraphics[width=10 cm]{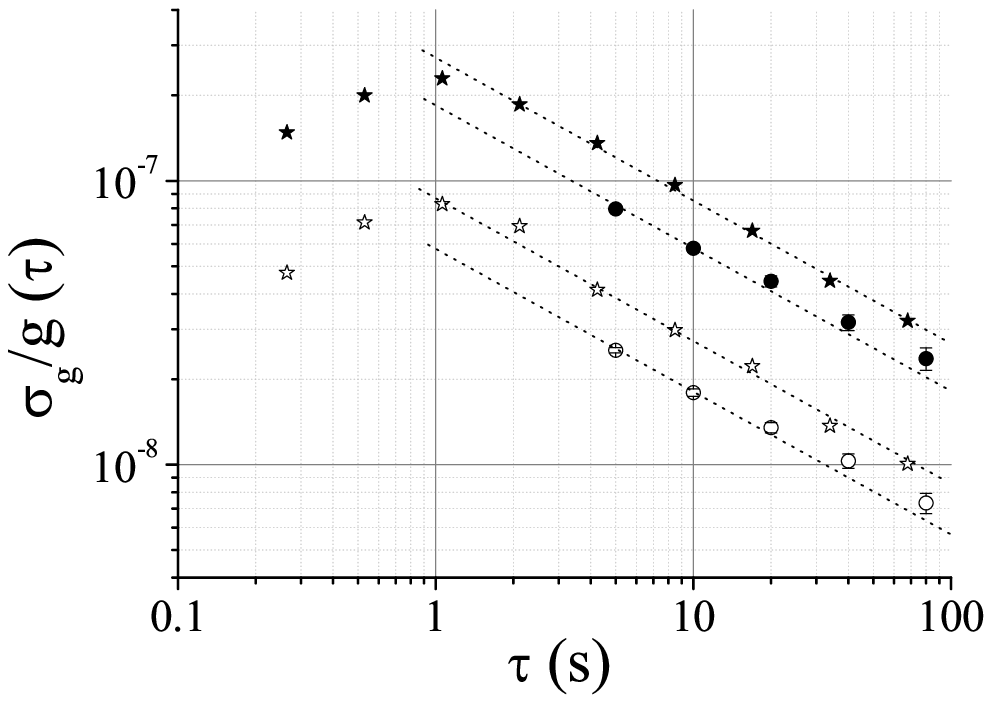}
    \caption{Allan standard deviation of g fluctuations versus averaging time. Measurements with the nonlinear lock technique at day (resp. at night) are displayed as full stars (resp. open stars). Measurements with the fringe fitting technique at day (resp. at night) are displayed as full circles (resp. open circles).}
    \label{best}
\end{figure}

Figure \ref{best} displays the Allan standard deviation of g fluctuations for $2T=100$ ms, with the two techniques described above (fringe fitting and nonlinear lock), during day and night times. The vibration phase shifts were calculated out of the 3D signals, using the optimal delay of 4.6 ms.
We obtain equivalent sensitivities at 1~s of $2.7\times10^{-7}$g (resp. $1.8\times10^{-7}$g) with the nonlinear lock (resp. fringe fitting) technique during the day, and $8.5\times10^{-8}$g (resp. $5.5\times10^{-8}$g) during the night.
We find that the fit of the fringes is slightly better than the lock technique, by about 50\%. The efficiency in removing vibration noise from the gravimeter signal can be calculated from the ratio of the sensitivities obtained here with the calculated contribution of the vibration noise (see section \ref{vibnoise}). A gain from 11 to 25 is obtained depending on the technique and noise conditions.

Best sensitivities are obtained during night measurements, as the vibration noise in the 1-10 Hz band is significantly lower. We reach at best an equivalent sensitivity as low as $5.5\times10^{-8}$g at 1~s when fitting fringes, which is only 4 times worse than our best reported value with the platform floating \cite{LeGouet08}, and only twice larger than the sensitivity obtained in our laboratory with a commercial FG-5 corner cube gravimeter \cite{Niebauer95} in the same vibration noise conditions.

\begin{figure}[h]
       \includegraphics[width=10 cm]{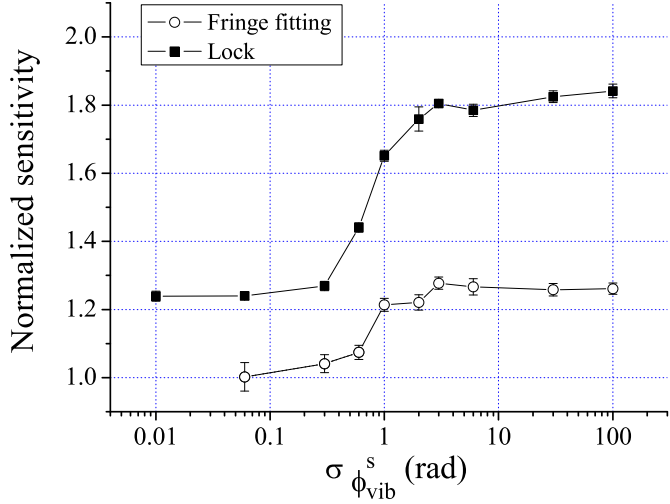}
    \caption{Numerical simulation of the normalized sensitivity of the interferometer as a function of the vibration noise standard deviation. Black squares (resp. open circles) display the sensitivity degradation for the nonlinear lock (resp. fringe fitting) technique.}
    \label{simu}
\end{figure}

These two techniques were also compared in a numerical simulation, where the phase of the interferometer was generated randomly as the sum of two independent terms $\phi=\phi_{1}+\phi_{2}$, with Gaussian distribution of standard deviations $\sigma_{1}$ and $\sigma_{2}$. $\phi_{1}$ simulates the vibration phase noise measured by the seismometer $\phi_{vib}^S$, and $\phi_{2}$ the phase difference between the real vibration phase noise and $\phi_{vib}^S$. We then implemented the two techniques with such simulated data, with $\sigma_{2}=0.02$ rad and with $\sigma_{1}$ ranging from 0.06 to 30 rad. For each technique, we find the corresponding sensitivity of the interferometer at 1 shot $\sigma_{\Phi}$ and calculate a normalized sensitivity by dividing $\sigma_{\Phi}$ with $\sigma_{2}$. We verified that this normalized sensitivity does not depend on $\sigma_{2}$. The results of the simulations are displayed in figure \ref{simu} and for both techniques the normalized sensitivity exhibits the same behavior. It increases for vibration noise larger than a few hundreds mrad, for which linear approximation of the transition probability is no longer valid, and finally saturates for large vibration noise. This degradation is due to the non linearity of the transition probability versus interferometer phase: measurements at top and bottom of the fringes have no sensitivity to phase fluctuations. The simulation confirms that this degradation is higher for the lock technique than for the fringe fitting technique, as observed in the measurements. In particular, for $\sigma_{2}=3$ rad, which corresponds roughly to day conditions, we find normalized sensitivities of 1.28 and 1.80 for the fringe fitting and lock techniques. The ratio of the sensitivities is thus 1.4, in reasonable agreement with the measurements.

\subsection{Investigation of systematic effects}

It is important to verify that the techniques presented here provide an accurate measurement of the interferometer phase, free from any bias. The lock procedure, which is intrinsically non-linear, could in principle induce such a bias. The numerical simulation indicates that none of the two techniques suffer from such systematics. This was confirmed experimentally by performing differential measurements, alternating the standard integration technique described in \ref{conv} with the lock procedure described in \ref{extension}, in the case where the platform was ON and thus the noise level low. The difference between the two techniques was found to be $0.3 \pm 0.8 \mu$Gal, which is consistent with no bias. Moreover, the two techniques were compared together during the day with the platform OFF, which corresponds to a noise level of $\sigma_{\phi_{vib}^S}=3$ rad. The difference for a 6 hours measurement was found to be $-2 \pm 4$ mrad, which corresponds to $-5\pm 10 \mu$Gal, which is also consistent with no bias.

\subsection{Interest of the nonlinear lock procedure}

\begin{figure}[h]
       \includegraphics[width=15 cm]{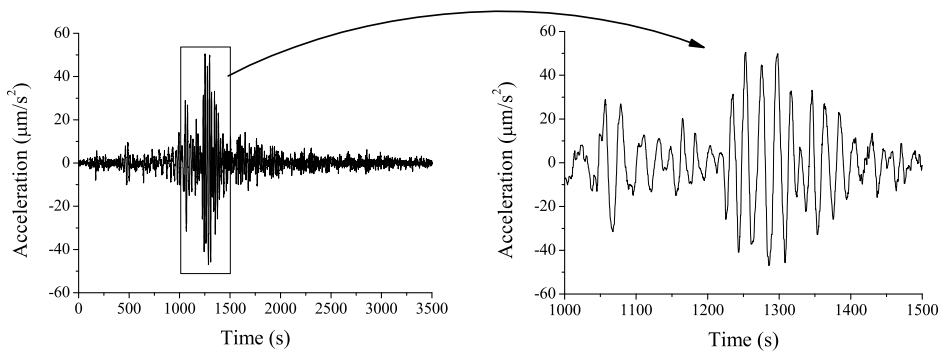}
    \caption{Fluctuations of the gravimeter signal during the earthquake of magnitude 7.7 that occurred in China on March 20$^{\textrm{th}}$ 2008. Data were obtained with the nonlinear lock procedure.}
    \label{seism}
\end{figure}

The main advantage of the nonlinear lock scheme is a better time resolution. Indeed, the time constant of the lock loop can be reduced to a few cycles only, so that a time constant $\leq 1$~s can be reached. In comparison, fitting the fringes requires to fit data by packets of at least 20 cycles for optimal sensitivity, which reduces the time resolution to about 5~s. Both techniques can operate with low vibration noise. Indeed, the fit of the fringes can also be adapted by modifying the phase modulation to add measurements performed at the top and bottom of the interferometer, in order to constrain the sinusoidal fit (doing so, sensitivity will as well be degraded because these measurements are not sensitive to phase fluctuations).
We finally illustrate the efficiency of the lock algorithm by demonstrating its robustness versus large changes in the vibration noise. Figure \ref{seism} displays the measurement during an earthquake of magnitude 7.7 that occurred in China on the 20th of March 2008. The gravimeter detects efficiently the occurrence of seismic waves, of period about 20 s. As our seismometer, of long period 30 s only, measures these vibrations with a large phase lag of about 1 radian, they are not efficiently removed from the gravimeter phase shift by the lock algorithm. They thus appear as a clear and well resolved signal in the gravimeter data. This demonstrates the robustness of our system versus large excitations, which is not the case for traditional absolute corner-cube gravimeters, which have neither adequate repetition rate (usually about 0.1 Hz) nor sufficient dynamic range, due to the finite range of the superspring mechanism. Note that the use of a longer period seismometer would in principle allow removing these low frequency vibrations from the gravimeter data.

\section{Conclusion}

In this paper, we demonstrate that an atom interferometer can reach high sensitivities without vibration isolation, when using an independent measurement of vibrations by a low noise seismometer. We develop here several measurement protocols that allow determining the mean phase of the interferometer, even when the interferometer phase noise amplitude exceeds $2\pi$. In particular, fitting the fringes scanned by vibration noise allows reaching a sensitivity as low as $5.5\times10^{-8}g$ at 1~s during night measurements. This performance is obtained with a rather short
interaction time ($2T=100$~ms), for which the vertical length of the interferometer
corresponds to a few centimeters only.

The techniques presented here are of interest for the realization of a portable atom gravimeter, with potential application to geophysics and gravity measurements in noisy environments. A compact gravimeter,
associated with a good AC accelerometer and operating at a high
repetition rate would reach fairly high
sensitivities, without much hardware isolation against ground
vibrations. Moreover, in contrast with other classical instruments, such as ballistic corner cube gravimeters, a high sensitivity would still be reached in the presence of earthquakes, if using a long period seismometer (100~s) to measure vibration noise.

More generally, these techniques can be extended to differential measurements with atom interferometers, such as gradiometers and cold atom gyroscopes. In particular, the phase difference can easily be extracted from the fits of the two interference patterns. Strong interests of these techniques lie in the ability of extending the dynamic range of the sensors, and of extracting the inertial phase without bias.

%\bibliography{apssamp}
{\bf Acknowledgments}

We would like to thank the Institut Francilien pour la Recherche sur les Atomes Froids (IFRAF) and the European
Union (FINAQS) for financial support. Q. B. and J. L. G.
respectively thank CNES and DGA for supporting their work.

\end{document}